\documentclass{elsart}
\begin{document}
\begin{frontmatter}
\title{Transitions between different superconducting states in
mesoscopic disks.} 
\author
{V.A. Schweigert \cite {*:gnu} and F.M. Peeters \cite {f:gnu}}
\address{\it  Departement Natuurkunde, Universiteit Antwerpen (UIA),\\
Universiteitsplein 1, B-2610 Antwerpen, Belgium}
\date{\today}
\maketitle
\begin{abstract}
Using a linear analysis, we study
the stability of  giant--vortex states in very thin disks.
The vortex 
expulsion and penetration fields are 
obtained for finite thickness disks
from a numerical solution of
the non-linear Ginzburg--Landau (GL) equations.
Using an extension of the London approximation, in which the phase
distribution of the order parameter is prescribed and the 
superconducting
electron density is found numerically,  we
consider the free energy behavior for transitions between different
superconducting states.
\end{abstract}



\end{frontmatter}
Recently, much attention is devoted  to investigations of
superconducting phenomena in mesoscopic disks whose radius $R$ and
thickness $d$ are comparable to the coherence $\xi$ and penetration
 $\lambda$
lengths. Precise
 measurements of the magnetization of single
superconducting $Al$ disks using
Hall-magnetometry allowed to follow the evolution of few vortex
states \cite{geim1,geim2}. A number of remarkable observations 
(the influence of
disk size on the {\it type} and {\it order} of the phase transition
 between
the normal and the superconducting state, prominent hysteretic
behavior in defect--free samples, the paramagnetic Meissner effect
in the field--cooling regime, etc.) initiated 
several theoretical works, in which superconductivity of mesoscopic
samples is considered within the  full GL approach.
The superconduncting state may present either a giant
vortex state with an axially symmetric distribution of the
superconducting  density 
or a multi-vortex Abrikosov-like
state, which is usually realized in type-$II$ superconductors.
Since the effective penetration length $\lambda_{\star}=\lambda^2/d$ 
\cite{gennes} may by far exceed that of a bulk superconductor, the
multi--vortex states can also exist in thin disks made from type-$I$
superconducting material.

The giant vortex states for finite thickness disks were considered  
in Refs.\cite{deo,schweigert}.
With increasing disk radius
a second-order reversible phase transition from the normal to the
superconducting state
observed for small disk radii is 
replaced by first-order transitions
with jumps in the magnetization \cite{deo,schweigert}.  
The simulated magnetizations are in  good 
quantitative agreement with those from experiment \cite{deo,deo1,deo2}.
The multi--vortex states have been investigated
using the London approach
\cite{buzdin}, the GL approach \cite{deo1,deo2,schweigert1}, and
the lowest Landau level approximation \cite{palacios}.
The free energy barriers separating the
superconducting states with different vorticity
were obtained using an expansion 
of the order parameter over
the eigenfunctions of the
 GL kinetic energy operator \cite{schweigert2}. 
The lowest Landau level approximation was used to study these
barriers in Ref.\cite{palacios1}.
An {\it equilibrium} phase diagram, showing which state is 
energetically preferable as function of the
magnetic field and disk thickness, was  found in 
Ref.\cite{schweigert1}. 
In the present work, we study a {\it non-equilibrium} phase diagram
showing the stability region of the giant vortex states as well
as the {\it order} of the transition between different
superconducting states.

We consider a superconducting disk surrounded by an insulator medium
and placed in a uniform magnetic field $H$, which is perpendicular to
the disk surface.
 Measuring the distance in units of the
coherence length $\xi$, the vector potential $\vec A$ in 
$c\hbar/2e\xi$,  and the order parameter $\Psi$
in $\sqrt{-\alpha/\beta}$ with $\alpha$, $\beta$ being the GL
coefficients \cite{gennes}, we write the system of GL equations for
a thin ($d\ll \xi,\lambda$) disk in
the following form \cite{schweigert1}:
\begin{equation}
\label{eq1}
\left(-i\vec \nabla_{2D} -\vec A\right)^2\Psi=
\Psi (1-|\Psi|^2),
\end{equation}
\begin{equation}
\label{eq2}
-\triangle_{3D} \vec A=\frac{d}{\kappa^2}\delta(z) \vec j_{2D},
\end{equation}
\begin{equation}
\label{eq3}
\quad \vec j_{2D}=\frac{1}{2i}
\left(\Psi^*\vec \nabla_{2D} \Psi-\Psi \vec\nabla_{2D}\Psi^*\right)
-|\Psi |^2\vec A,
\end{equation}
where the indices $2D$, $3D$
refer to two-dimensional (in the disk plane $(x,y)$)
and three-dimensional operators;
$\kappa=\lambda/\xi$ is the GL parameter;
$\vec j_{2D}$ is the density of 
superconducting current. 
The boundary conditions to Eqs.~(\ref{eq1},\ref{eq2}) correspond to
zero superconducting current at the disk boundaries and an uniform
external magnetic field far from the disk 
$\vec A_{|\vec r|\rightarrow \infty} =
\frac{1}{2}[\vec r\times \vec H_0]$, respectively. Our numerical
approach to solve Eqs.(\ref{eq1},\ref{eq2}) is described in 
\cite{schweigert1}.

We restrict our considerations of the stability of the giant vortex
 states
to the case of small disks $\lambda_{\star}\gg R$,
when the expulsion of the magnetic field can be neglected. Then the
order parameter
of the giant vortex state can be presented 
as $\Psi=\psi(\rho)e^{iL\phi}$,
where $L$ is the angular quantum momentum; $\rho$, $\phi$ are the
cylindrical coordinates. The radial wavefunction $\psi$
 obeys the following equation
\begin{equation}
\label{eq5}
-\frac{1}{\rho}\frac{\partial }{\partial \rho}\rho
\frac{\partial \psi }{\partial \rho}+(\frac{L}{\rho}+
\frac{1}{2}H_0\rho)^2\psi=\psi-\psi^3,
\end{equation}
with the boundary condition $(\partial \psi/\partial \rho)_{\rho=R}=0$.
In order to investigate whether 
the giant vortex state is stable with respect to small
perturbations from  other angular momentum states we use the
time-dependent (TD) first GL equation \cite{hu}.  Representing
the order parameter as a mixture of three angular harmonics
$\Psi=\psi (\rho) e^{iL\phi}
+\delta_1(\rho) e^{iL_1\phi+\mu t}+\delta_2(\rho) e^{iL_2\phi+\mu t}$,
we linearize the first TDGL equation with respect to small
 perturbations
$\delta_1$, $\delta_2$ and obtain the following set of equations
\begin{equation}
\label{matr}
\left |
\begin{array}{lr}
\hat G_{L1}+1-2\psi^2&-\psi^2\\
-\psi^2&\hat G_{L2}+1-2\psi^2 
\end{array}
\right |\cdot
\left |
\begin{array}{c}
\delta_1\\
\delta_2 
\end{array}
\right |=\mu
\left |
\begin{array}{c}
\delta_1\\
\delta_2
\end{array}
\right |,
\end{equation}
which describes our eigen-value problem. Here,
$$\hat G_{L}=-\frac{1}{\rho}\frac{\partial }{\partial \rho}\rho
\frac{\partial }{\partial \rho}+(\frac{L}{\rho}+
\frac{1}{2}H_0\rho)^,2$$
is the Schr\"odinger operator in an uniform magnetic field 
with the boundary condition 
$(\partial \delta_{1,2}/\partial \rho)_{\rho=R}=0$.
The sign of the eigenvalue $\mu$ determines whether
the considered state is stable $\mu<0$ or unstable $\mu>0$.
Note, that the use of three harmonics
and the relation $L_1+L_2=2L$ between them is dictated by the
non-linear term in the RHS of Eq.(\ref{eq1}), which corresponds to the
third power of the order parameter. Generally, a linear analysis 
does not allow to predict which state will be realized from 
the evolution of instability. However, according to 
our previous results from solving the non-linear
GL equations \cite{schweigert1} we expect that the
perturbation with $L_1=0$, $L_2=2L$ leaves the vorticity unchanged
and leads to the appearance of a state corresponding
to $L$ vortices arranged in a ring.
When the angular momentum of the perturbation is not a multiple of that
 of
the initial state, we  expect the appearance of a
supercondicting state with a different vorticity.
To find the
perturbation spectrum we use a
finite-difference representation of the Schr\"odinger operator $\hat G$
\cite{schweigert} 
and reduce Eq.(\ref{matr})
to a matrix, which is numerically diagonalized with the
Householder technique. The state becomes unstable if 
the maximum value of $\mu$ found for different pairs $L_1,L_2=2L-L_1$
changes its sign and becomes positive. 

The results from our linear analysis 
are shown in Fig.~\ref{fig1} for $L=0,...,9$.
The Meissner state becomes unstable relative 
to the entrance of extra vortices with
increasing magnetic field above the penetration field $H_p$.
For small
disk radii, it is the perturbation with $L_1=L-1,L_2=L+1$ which makes
the $L$ state unstable.
With further increasing the disk radius, our linear analysis shows that
a couple of vortices will enter the system at once.
The points, at which the number of penetrating vortex changes,
are denoted by solid squares in Fig.~\ref{fig1}.
A similar behavior of
vortex penetration is also observed for the single vortex state.
 It should be noted, that the eigenvalues of the perturbations
with different angular momenta are very close to each other at large
disk radii. Therefore, a non-linear
consideration is required to answer how many vortices can 
simultaneously enter
 into the disk. Unfortunately, such an analysis
is a very difficult task since the results turn out to be very 
sensitive to the
shape of the
perturbation. Our numerical solution of 
the non-linear GL equations 
shows that occasionally several vortices enter 
indeed into the system.
Although we are able to 
predict the entrance field 
 with  high accuracy,
the number of penetrating vortices is rather uncertain.
This indicates the at the entrance field the barriers separating the
different $L$-states become very small.

The Meissner state and the single vortex state can exist in arbitrary
large disks. The expulsion field $H_e$, below which the vortex state
becomes unstable relative to the transition to the Meissner state,
decreases with disk radius, in such a way that 
the external magnetic flux piercing the disk tends to the flux quantum
$\Phi_0=hc/2e$.
The other giant vortex states remain stable only inside some
region of the phase diagram.
 Their stability is
restricted by either the transition to the multi--vortex state
with the same vorticity, which occurs as a rule
 with decreasing magnetic field,
or the transition to another
giant vortex state. The left (right) boundary of
the stability region corresponds to expulsion (penetration) of a
 single
vortex with decreasing (increasing) disk radius. 
 Note, that the state with $L=2$
in principle allows for a reentrant behavior - the transition to
the multi--vortex state with both decreasing and increasing magnetic
 field.
Unfortunately, this prediction is not confirmed by our non-linear
analysis. The boundary of transition between the giant and
multi--vortex state increases with  angular momentum and,
as expected, tends to saturate at the
second critical field $H_{c2}$ for large disk radii. The stability
regions of different states overlap strongly
and many different superconducting
states can exist with the same external conditions.

To study the influence of the disk thickness on the penetration and
expulsion field we solve numerically Eqs.~(\ref{eq1},\ref{eq2}).
Starting from the Meissner (vortex) state we increase (decrease)
slowly the applied magnetic field.
 Using the superconducting state obtained at the
previous step as initial condition, 
we  find a new steady-state solution of
Eqs.~(\ref{eq1},\ref{eq2}) and check whether the vorticity changes.
The penetration and expulsion fields found for zero disk thickness
coincide with those from the linear analysis above. It turns out that
the vortex state exibits a weak paramagnetic Meissner effect just
before vortex expulsion. This is the main reason why the
expulsion field (thin dotted curves in Fig.~\ref{fig2}) decreases
with disk thickness. The Meissner state shows, as expected, a strong
diamagnetic response. The decrease of the total magnetic flux
piercing through the disk  results in
an expansion of the stability region of the Meissner state
with increasing $d/\lambda^2$.
Recall, that the field expulsion from thin disks is only determined
by  the ratio $d/\lambda^2$, which
accounts for both
the disk thickness and the penetration length of bulk material.
The first critical field $H_{c1}$, above
which the vortex state becomes energetically more favorable, 
also increases with the disk thickness.
All the critical fields $H_{c1}=2H_{c2}\xi/R$, 
$H_e=2H_{c2}\xi^2/R^2$, and $H_p=2H_{c2}\xi ln(R/\xi)/R$ 
can be found within the
London approximation in the limit of zero disk thickness
\cite{buzdin,schweigert2}.
Note, that 
the penetration field from the London approximation does not
agree with that from the GL theory because the saddle point state
for vortex entrance does not correspond to any vortex state
nearby $H_p$ \cite{schweigert2}.
 However, the London approach
predicts rather accurately the expulsion and the first critical field
 for
$R\gg \xi$ disk radii.

The superconducting  density is assumed to be a constant
in the conventional London approximation. This assumption breaks
down at small disk radii or small inter-vortex distances which is
 the case
of mesoscopic samples. An extension of the London approach accounting
for variation of the amplitude $\psi$ of the order parameter
$\Psi(\vec \rho)=\psi(\vec \rho)e^{iS(\vec \rho)}$ has been proposed in
\cite{schweigert2}.  
The phase distribution $S$, which depends on the vortex positions
$\vec \rho_i$, is assumed to be created by the vortices and their 
mirror images with coordinates $\vec \rho_i$ and $\vec \rho_i
R^2/\rho_i^2$, respectively \cite{buzdin,schweigert2}. Then the
amplitude of the order parameter obeys the first GL equation
\begin{equation}
\label{final}
-\triangle \psi+(\vec\nabla S-\vec A)^2\psi
=\psi-\psi^3,
\end{equation}
with the boundary condition $(\partial \psi/\partial \rho)_{\rho=R}=0$.
Solving numerically the last equation we find the free energy
$F(\vec \rho_i)=-\int \psi^4 d\vec r/\pi dR^2$, where integration is
 performed
over the disk volume. Doing so
we reduce the GL free energy functional
defined in the functional space to the functional of the vortex 
coordinates.
Then the vortices can be considered as classical particles,
whose motion is governed by this free energy. Such an approach, which
we call the modified London approach below, is
shown to lead to accurate results in the case when vortices do not 
cross
the sample boundary \cite{schweigert2}.

According to numerical simulations \cite{schweigert1}, the transition
from the giant vortex state to the multi--vortex state with the same
 vorticity
is followed by a weak jump in the slope of the 
magnetization indicating
a second--order transition. The modified London approach
allows to study the free energy as function of the vortex
coordinates, which is shown in Fig.~\ref{fig3}
for the two--vortex state. 
At large magnetic fields,  the minimum of the free
energy is achieved when both vortices are located in the disk center.
With decreasing  magnetic field, the curvature of the potential
curve decreases and tends to zero  at some critical point
$H/H_{c2}\approx 0.58$. Thereafter, the multi--vortex state with 
separated
vortices becomes more energetically favorable. No
two minima of the free energy exist simultaneously. This agrees
with results from our numerical solution of the
 GL equations \cite{schweigert1}
and proves that the transition is of second--order.

At small disk radii, all the multi--vortexs state correspond to
vortices arranged in a ring. 
With increasing disk radius, the configuration $(1:L-1)$
with a single vortex inside the ring 
also becomes stable at large vorticity and
coexists with the ring structure $(0:L)$ 
\cite{schweigert3}.
There is experimental evidence of such a 
coexistence \cite{geim3}.
Solving numerically Eqs.~({\ref{eq1},\ref{eq2}) we find the
magnetization of both state $(0:7)$, $(1:6)$ (see Fig.~\ref{fig4}).
For $R=5\xi$, the state with 
$(0:7)$ turns out to be energetically preferable.
 Note, that the magnetization curves
do not merge with each other suggesting a first--order 
transition between these multi--vortex states.
To find the free energy barrier
separating the states $(1:6)$, $(0:7)$ we apply the modified London
approach. Starting from the configuration $(1:6)$ we move slowly 
the central
vortex to the disk boundary. The coordinates of
the other vortices are found by minimizing the free energy. 
At small magnetic fields, there exists
two minima of the free energy separated by a barrier
 (see Fig.~\ref{fig5}).
 At some
critical field $H/H_{c2}\approx 0.82$ this barrier
disappears and the state $(1:6)$ becomes unstable. The
transition between states ($1:6$) and ($0:7$) is followed by a jump
in both the magnetization and the free energy which also proves 
the first--order of the transition between the multi-vortex states.
Note, that the barrier separating two multi--vortex states is much
smaller than that for vortex expulsion and penetration.

 We thank A.K.~Geim for
usefull discussions.
This work is supported by the Flemish Science Foundation (FWO-Vl)
 and the
``Interuniversity Poles of Attraction Program - Belgian State,
 Prime Minister's
Office - Federal Office
 for Scientific, Technical and Cultural Affairs''.
One of us (FMP) is a research director with the FWO-Vl.

\newpage

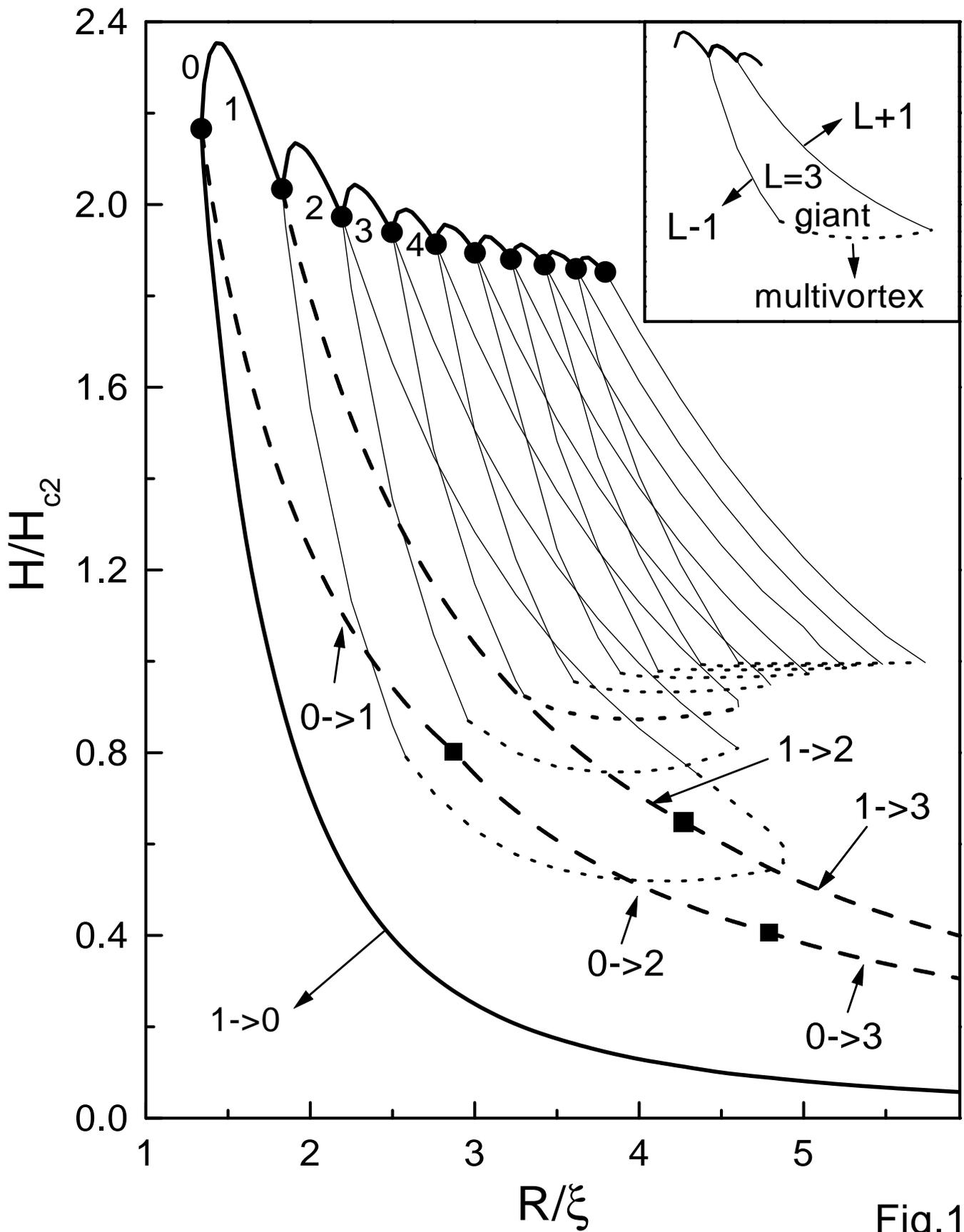
\begin{figure}
\caption{
The stability region of the different giant vortex states for zero
disk thickness. The upper and lower solid thick curves correspond
to the nucleation $(H_{c3})$ and expulsion $(H_e)$ field, respectively.
The thick--dashed curves
denote the penetration fields $(H_p)$ for the Meissner state and
the single vortex state.
The thin solid and dotted curves correspond to the transitions
between the giant vortex states with different angular momenta and
the transition from the giant vortex state to the multi--vortex state
with the same vorticity (see inset) , respectively.
}
\label{fig1}
\end{figure}

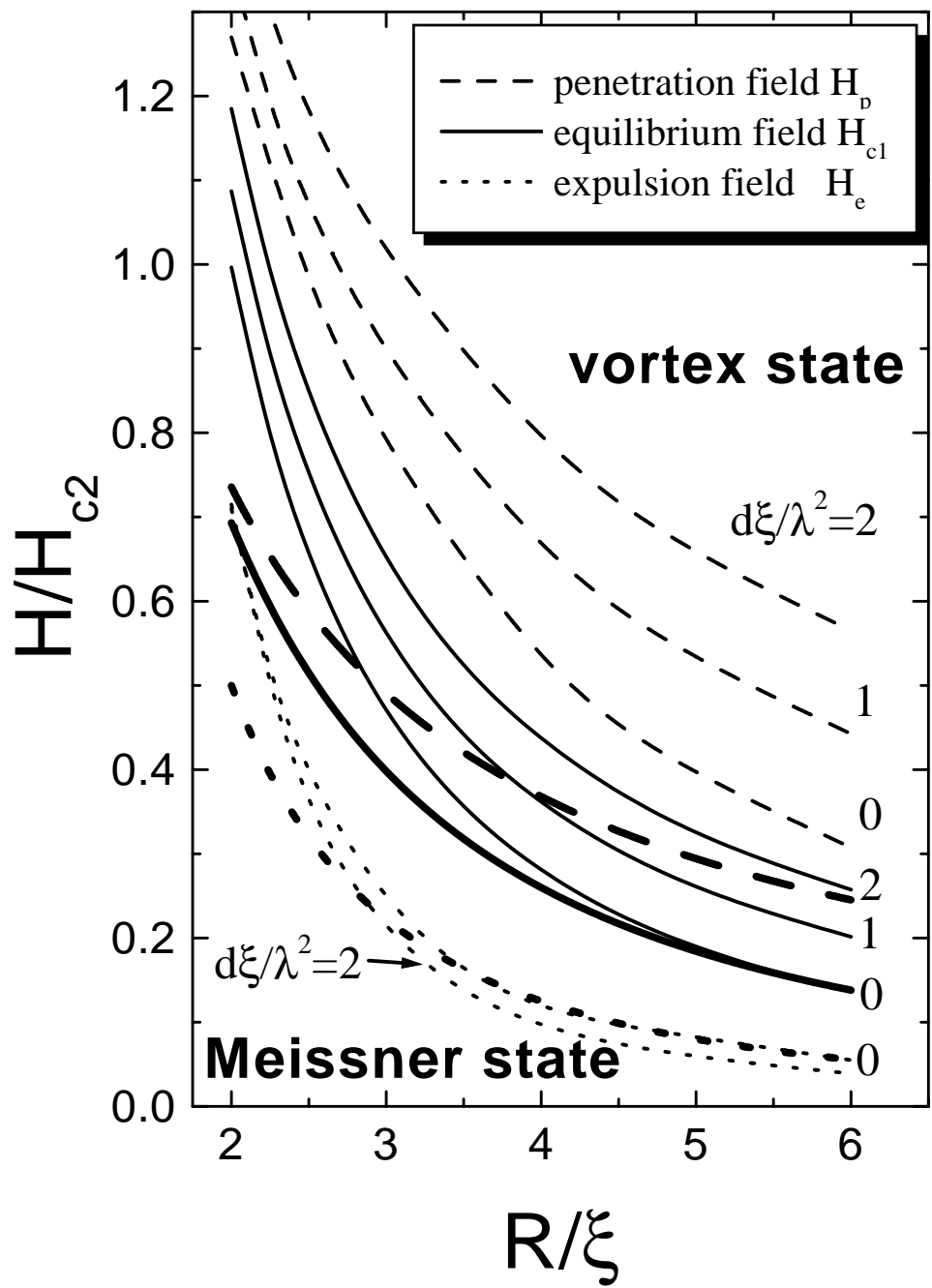
\begin{figure}
\caption{
The first critical ($H_{c1}$), penetration ($H_p$), and 
expulsion ($H_p$)
field as function of the disk radius for different disk thicknesses.
Thick curves denote results found with a conventional London approach.
}
\label{fig2}
\end{figure}

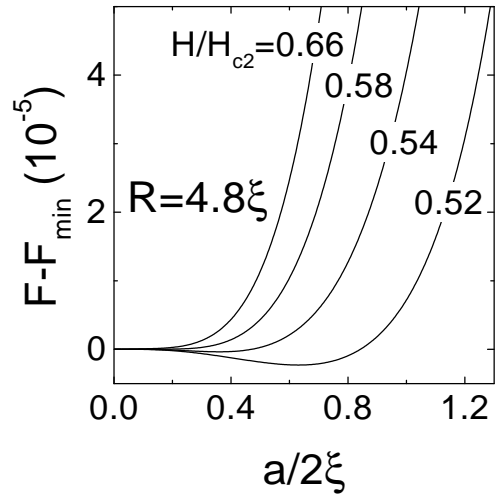
\begin{figure}
\caption{
The free energy of the two-vortex state as a function of the
 inter-vortex
distance (a) for different magetic fields.
}
\label{fig3}
\end{figure}

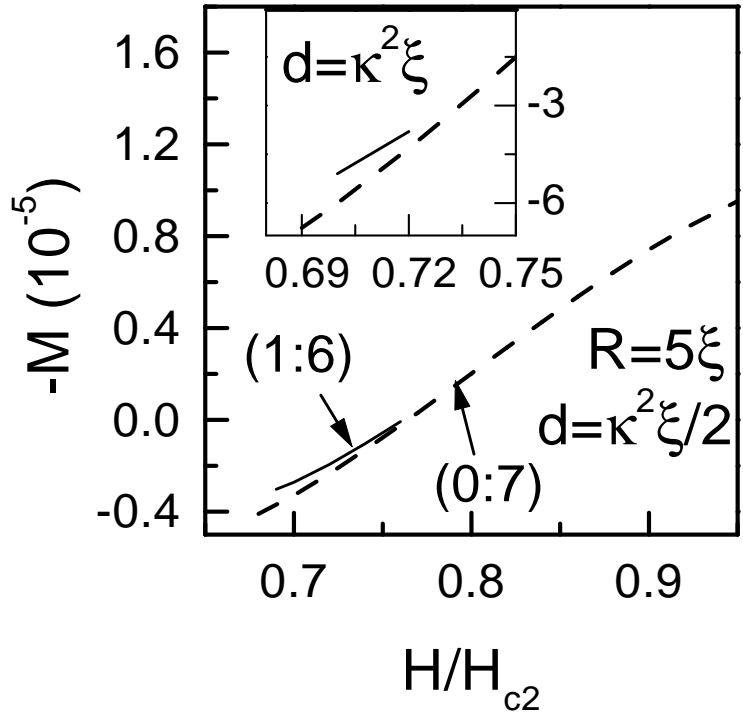
\begin{figure}
\caption{
The unitless magnetization 
$M=\int (H-H_0)d\vec \rho /4\pi^2 H_{c2}R^2$ of the multivortex
state $L=7$ for two different vortex configurations and two
thicknesses of the disk.
The calculated
magnetization includes the detector size effect \protect\cite{deo2}
(square detector with width $\approx 3.1\mu m$).
}
\label{fig4}
\end{figure}

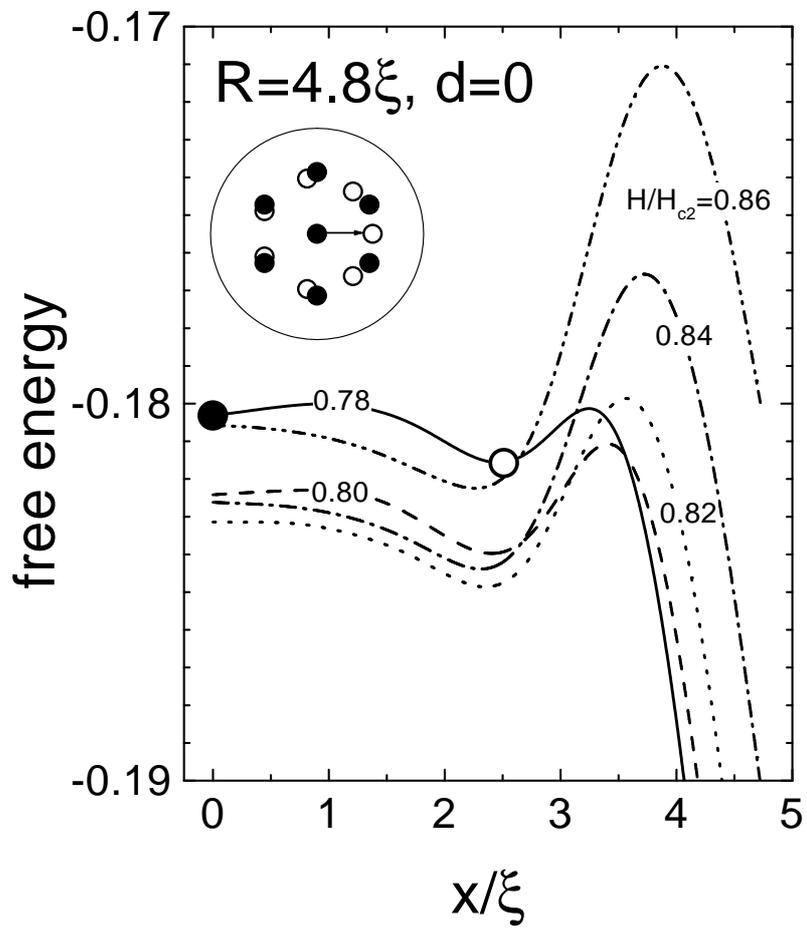
\begin{figure}
\caption{
The free energy as a function of the position of one of the vortices,
 which is
shifted from the disk center to the disk boundary. The  
inset shows the vortex configurations in the two stable states.
}
\label{fig5}
\end{figure}


\begin{thebibliography}{BM}
\bibitem{schweigert3}
Our studies of large disks $R/\xi=8$ show the existence of 
a number of different stable 
configurations of the multi--vortex state with  large vorticity.
Similar to clusters of classical particles \cite{bedanov}, 
these configurations have
a shell structure with different shell occupation numbers.
\end{thebibliography}
\end{document}